# Flexible Hall Sensors Based on Graphene

Zhenxing Wang,[*] Mehrdad Shaygan, Martin Otto, Daniel Schall, Daniel Neumaier

Advanced Microelectronic Center Aachen (AMICA), AMO GmbH, Otto-Blumenthal-Straße 25, 52074 Aachen, Germany

[*]Author to whom correspondence should be addressed. Email: wang@amo.de

Abstract

The excellent electronic and mechanical properties of graphene provide a perfect basis for high performance flexible electronic and sensor devices. Here, we present the fabrication and characterization of flexible graphene based Hall sensors. The Hall sensors are fabricated on 50 µm thick flexible Kapton foil using large scale graphene grown by chemical vapor deposition technique on copper foil. Voltage and current normalized sensitivities of up to 0.096 V/VT and 79 V/AT were measured, respectively. These values are comparable to the sensitivity of rigid silicon based Hall sensors and are the highest values reported so far for any flexible Hall sensor devices. The sensitivity of the Hall sensor shows no degradation after being bent to a minimum radius of 4 mm, which corresponds to a tensile strain of 0.6%, and after 1,000 bending cycles to a radius of 5 mm.

**Introduction**

Hall sensor devices are widely used magnetic field sensors in automotive and consumer electronic applications, where they precisely detect position or monitor current.[1-4] The sensitivity, as one of the key parameters for Hall sensors, depends on the charge carrier mobility $\mu$ and the sheet carrier density $n$. Higher $\mu$ together with lower $n$ results in larger sensitivity.[1-3] Currently, silicon CMOS based Hall sensors are used in most of the applications, mainly due to their well-developed and low-cost production.[3-6] However, several emerging areas of application such as wearable electronics, electronic skin, highly integrated machines, robotics or prosthesis require thin and flexible sensor devices.[7-17] For those applications, flexibility provides not only an additional feature during operation, but the thin body also enables a more simple integration process on curved and non-planar surfaces. Therefore other approaches are currently explored to fabricate thin and flexible Hall sensors. In a recent work, a flexible and only 100 μm thick Hall sensor has been demonstrated using a thin bismuth layer.[17] Despite the good flexibility and a thickness of only 100 μm, the sensitivity of the device was about a factor of 40 lower compared to silicon CMOS based Hall sensors, which is not sufficient for many applications.[3,5] Graphene, the two-dimensional carbon allotrope, offers very high carrier mobility for both rigid and flexible substrates, low sheet carrier density and excellent flexibility, hence it can be considered as the perfect material for Hall sensors in general and for flexible Hall sensors in particular.[18-20] So far, different graphene based Hall sensors exploiting the excellent electronic properties have been realized on rigid silicon based substrates, however the excellent flexibility of graphene has not been utilized in these devices.[21-23] Here, we present the fabrication and characterization of thin and flexible Hall sensors based on graphene that show a sensitivity significantly above that of state-of-the-art flexible Hall sensors and similar to silicon CMOS based Hall sensors in combination with excellent flexibility.

**Results and Discussion**

The graphene based Hall sensors presented in this work were fabricated on 50 μm thick flexible Kapton foil using conventional thin film processing techniques. For sample handling during device fabrication the Kapton foil was fixed on a silicon chip using polydimethylsiloxane (PDMS). The adhesion between PDMS and Kapton foil is strong enough to enable the sample to bear all the wet fabrication processes, such as sample cleaning, development and lift-off.[19] To smoothen the Kapton surface, 600 nm of SU-8 photoresist was spin-coated and hard-baked. Subsequently commercial graphene grown by chemical vapor deposition on copper foil (Graphenea SE) was transferred using the standard wet-chemical release process and PMMA as a support layer.[24] The whole device fabrication was carried out by contact photo-lithography using AZ5214E as resist. The graphene was first contacted with 50 nm nickel using sputtering deposition and lift-off technique, and then patterned by oxygen plasma. A schematic of the Hall device is shown in Figure 1a. After fabrication, the complete flexible stack was peeled off mechanically from the PDMS surface in order to test the bending performance as shown in Figure 1b. An optical image of the final device is shown in Figure 1c and an optical micro-graph of one Hall cross is shown in Figure 1d. The distance between two electrodes is 18 $\mu$m and the width of graphene strip is 13 $\mu$m.

The measurement setup for testing the Hall sensor is depicted in Figure 2a. All measurements were performed on a probe station at room temperature under ambient condition using a HP 4156B semiconductor parameter analyzer. The magnetic field was generated by an electromagnet and the magnetic field at the sample position was calibrated using a commercial Hall sensor (Allegro MicroSystems A1324). The two-terminal *I-V* measurement of one graphene Hall sensor gave a constant current flow $I_C$ of 62 μA at a bias voltage $V_C$ of 50 mV, corresponding to a two-terminal resistance of about 810 Ω. This resistance value corresponds to a sheet resistivity $R_\square = 600$ Ω, including contact resistance, which is a typical value for CVD grown graphene without any intentionally doping steps. The measured Hall voltage $V_H$ responding to the perpendicular applied magnetic field ranging from -18 mT and 18 mT is shown in Figure 2b, with the bias voltage fixed at 50 mV and a constant current flow of 62 µA. The magnetic field was intentionally brought back to zero each time after a specific magnetic field was applied, in order to exclude any possible drift of the Hall voltage offset. The Hall voltages under different magnetic fields were extracted based on an average over time, and they are plotted as a function of magnetic field in Figure 2c, showing a linear dependence on the applied magnetic field. The voltage normalized sensitivity

$$S_V = \frac{1}{V_C}\left|\frac{\partial V_H}{\partial B}\right| \qquad (1)$$

and current normalized sensitivity

$$S_I = \frac{1}{I_C}\left|\frac{\partial V_H}{\partial B}\right| \qquad (2)$$

of the Hall sensor can be calculated based on the slope of the line plotted in Figure 2c.[1,2] A linear fit to the experimental data gives a slope of 4.64 mV/T, and hence $S_V$ = 0.093 V/VT and $S_I$ = 75 V/AT are extracted. The typically measured sensitivity values for Hall sensors based on different materials and substrates are compared in Table I. The sensitivity of the flexible Hall sensor in this work is already matching that of rigid silicon based Hall sensors, which are the most broadly used Hall sensors. Furthermore, compared to state-of-the-art flexible Hall sensors, the graphene Hall sensor presented here provides a factor of 30 higher sensitivity.[17] The highest sensitivity values for Hall sensors so far were obtained using high quality graphene embedded in hexagonal boron nitride (hBN), with a sensitivity of 1 to 2 orders of magnitude higher compared to this work.[23] Although these values provide an outlook on the ultimate sensitivity potential also for flexible graphene Hall sensors, it shall be noted that these record values have been obtained on rigid silicon substrates using non-scalable micromechanical exfoliated graphene and hBN.

Two important quantities to qualify the graphene layer are the carrier mobility $\mu$ and the sheet carrier density $n$, which can be derived from the voltage and current normalized sensitivities by:

$$\mu = S_V \frac{L}{W} \tag{3}$$

and

$$n = \frac{1}{S_I e} \tag{4}$$

respectively, where $W$ and $L$ are the width and length of the Hall sensor device, and $e$ is the elementary charge.[1,2] This gives a mobility for the graphene of 1300 cm$^2$/V·s, and a carrier density of $8.3 \times 10^{12}$ cm$^{-2}$. While the mobility is typical for CVD grown graphene on flexible substrates, the relatively high doping is related to the fact that the devices were not encapsulated and measurements were performed under ambient conditions. Therefore device encapsulation is expected to significantly reduce the doping level and hence increase the current normalized sensitivity.[25] However, depending on the flexibility of the encapsulation layer, this improvement might come at the cost of a reduced reproducibility after device bending.

The bias dependence of the sensitivities was also investigated on the same device, and they are plotted in Figure 2d. With a varying bias voltage from 10 mV to 200 mV, the current normalized sensitivity $S_I$ shows an average at 72 V/AT with a standard deviation of 3 V/AT, and the voltage normalized sensitivity $S_V$ shows an average at 0.090 V/VT with a standard deviation of 0.003 V/VT. From the small values of the standard deviation, we conclude that the Hall sensors are able to work very stable under different applied bias conditions. The application of bias voltages of 500 mV and larger leads to drift of the device resistance, which can be explained by self-heating and the corresponding desorption / adsorption of adsorbates from the graphene surface. Stable operation was not possible anymore under these high bias conditions.

To evaluate the uniformity of the device performance, we measured all working 27 Hall sensors out of a single sample and the histograms for the sensitivities are shown in Figure 3. All the measurements were carried out with an identical bias voltage of 100 mV and a magnetic field ranging from -18 mT to 18 mT for Hall measurements. The sensitivities were extracted based on the slope of the $V_H$-$B$ relation. The voltage normalized sensitivity $S_V$ is ranging from 0.042 to 0.096 V/VT, with an average value of 0.065 V/VT. The current normalized sensitivity $S_I$ is ranging from 33 to 79 V/AT, with an average of 49 V/AT. The distribution of the sensitivity can be attributed to the quality variation of CVD graphene in terms of mobility and doping level, and we note that the variation of the sensitivity is similar to what is achieved for graphene Hall sensors on rigid silicon substrate.[21]

Finally, bending tests were performed on the flexible substrate, by placing the sample onto the surface of different rod molds with different cross-section diameters, so the bending radius simply follows the dimension of the molds and can be easily determined. The Kapton foil was bent under different bending radii of 14 mm, 8 mm, and 4 mm, in sequence and for each radius one time bending, with the graphene based Hall sensor being on the outer site and therefore exposed to tensile strain. The tensile strain can be estimated from $T/2R$, where $T$ is the thickness of the Kapton substrate, i.e. 50 μm, and $R$ is the bending radius.[26] For a bending radius of 4 mm, the tensile strain applied is approximately 0.6%. After being bent under a specific strain, the sensors were measured in flat status again. Since the current mode is most widely used for conventional Hall sensors, we compare the current normalized sensitivity $S_I$ before and after the bending in Figure 4a.

Additionally bending cycles up to 1000 times were performed for one Hall sensor. Each bending cycle includes bending with radius of 5 mm and releasing of the substrate. The measurement of the sensitivity was done with the substrate released after a certain bending cycles. The evolution of the voltage and current normalized sensitivity is shown in Figure 4b, with values of 0.045 V/VT and 62 V/AT before the bending cycles and values of 0.039 V/VT and 65 V/AT after 1000 bending cycles. On the same device, the sensor function was also tested while bending with a radius of 5 mm. The voltage normalized sensitivity was 0.040 V/VT and the current normalized sensitivity was 66 V/AT for the sensor in the bent status, which are similar values to that at flat status 0.039 V/VT and 65 V/AT, respectively. The sensitivity of the Hall sensors shows no degradation after and during bending, demonstrating the full flexibility of the graphene based Hall sensor devices.

**Conclusions**

In summary, flexible Hall sensors based on graphene are realized, which show a voltage and current normalized sensitivity of up to 0.096 V/VT and 79 V/AT, respectively, comparable to rigid silicon based Hall sensors and significantly outperforming all flexible Hall sensors so far by more than one order of magnitude. A further increase of the sensitivity of graphene based Hall sensors is still possible using either a flexible encapsulation or ultimately high quality graphene encapsulated in hBN. The graphene Hall sensors are only 50 μm thick and the sensitivity is stable for bending radius down to 4 mm and for 1,000 bending cycles to a radius of 5 mm. Therefore graphene based Hall sensors will enable upcoming applications in the fields such as bio-medicine, wearable electronics or electronic skin, where precise position detection or current monitoring is required.

**Acknowledgements**

This work was financially supported by the European Commission under the projects Graphene Flagship (contract no. 604391), SPINOGRAPH (contract no. 607904) and by the German Science Foundation (DFG) within the priority program 1796 FFlexCom Project "GLECS" (contract no. NE1633/3).

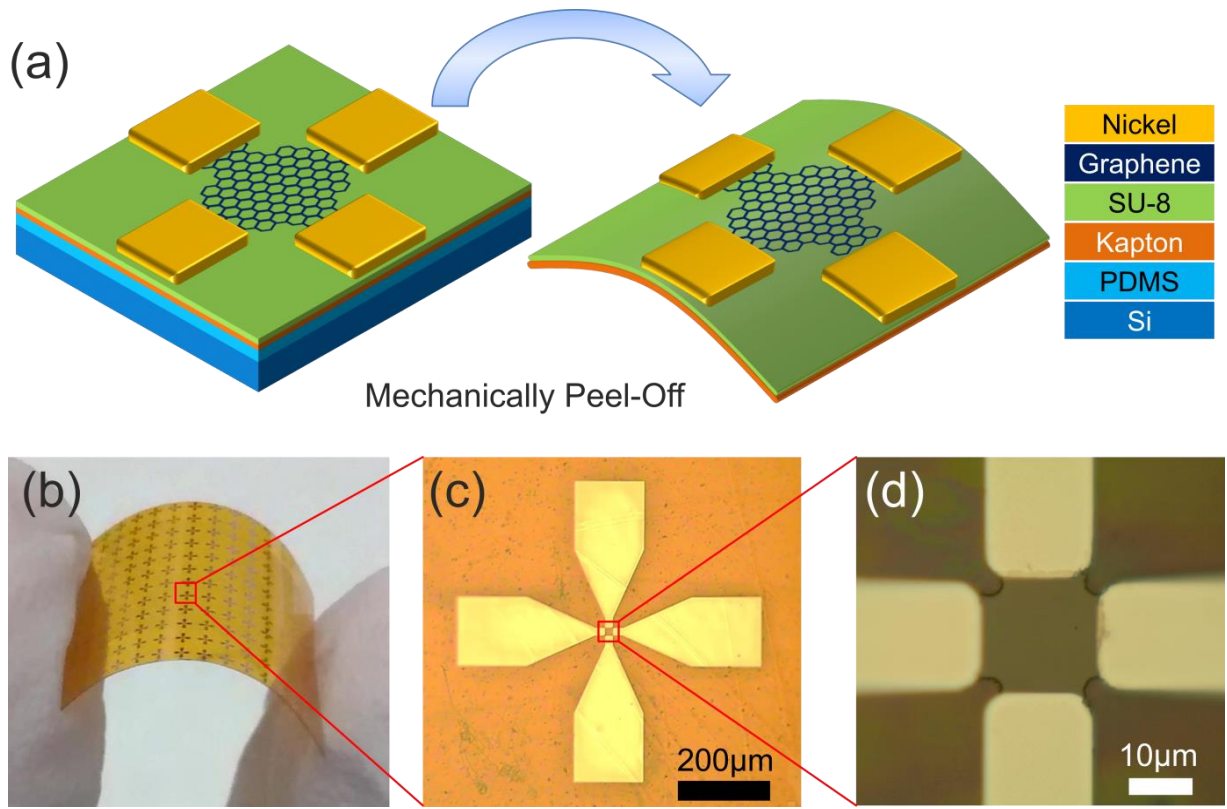

Figure 1. The device architecture and the geometry of the flexible Hall sensor based on graphene. (a) The device stack is fixed on a rigid substrate, i.e. silicon chip, for sample handling during processing. After fabrication, the whole flexible stack is peeled off from the silicon chip. (b) Optical photograph of the flexible chip with an array of Hall sensors. (c) An optical image of a single Hall sensor with four probing metal pads. (d) Detailed view of the core area of the Hall sensor. The distance between two electrodes is 18 $\mu$m and the width of graphene strip is 13 $\mu$m.

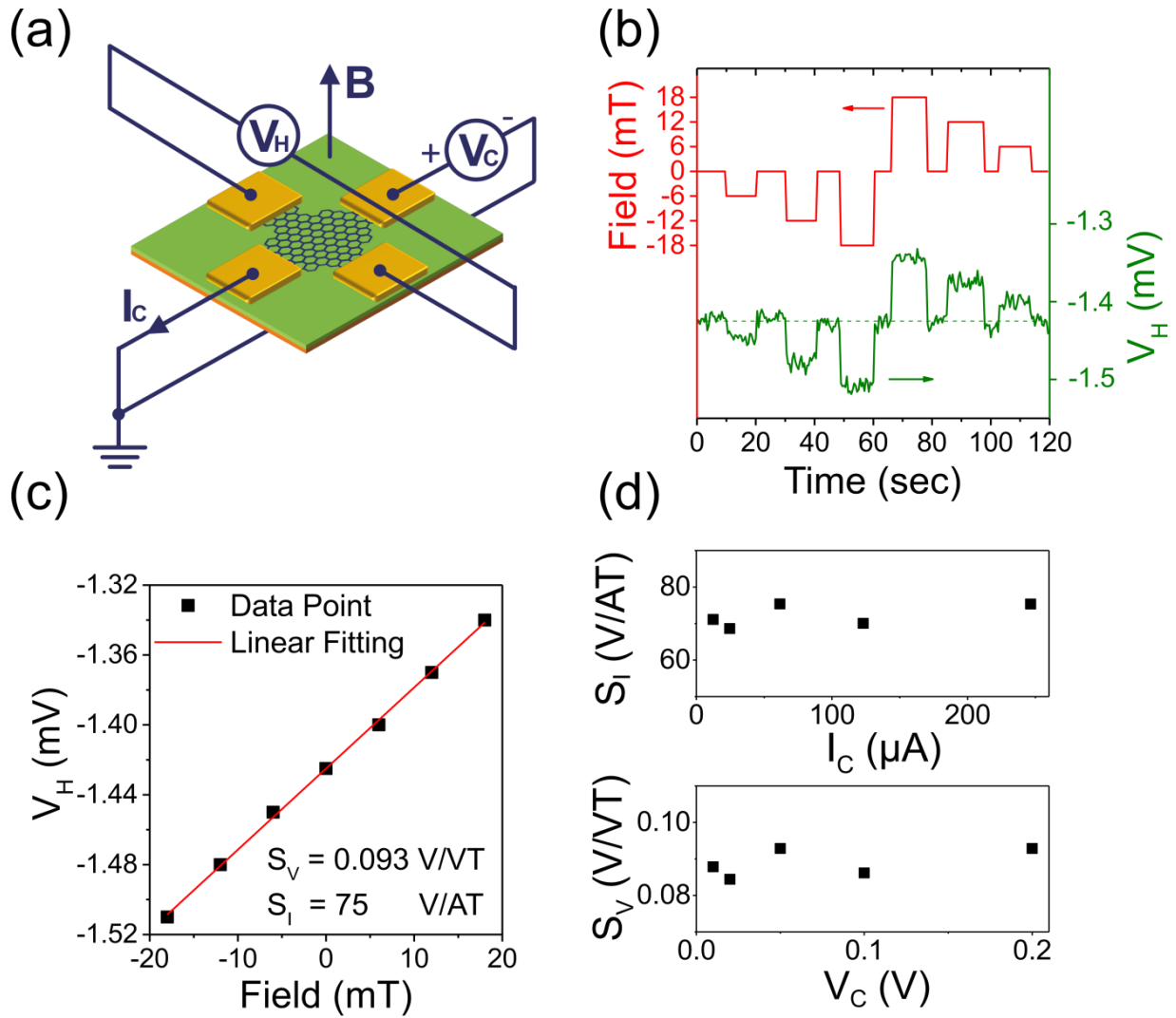

Figure 2. The Hall-effect measurements of the sensor. (a) Four terminal Hall-measurement configuration of the Hall sensor. (b) Response of the Hall voltage to the varying magnetic field. (c) The linear relationship between the Hall voltage and external magnetic field. The data points are extracted from (b) with an average on time. (d) The bias current dependence of the current normalized sensitivity and the bias voltage dependence of the voltage normalized sensitivity.

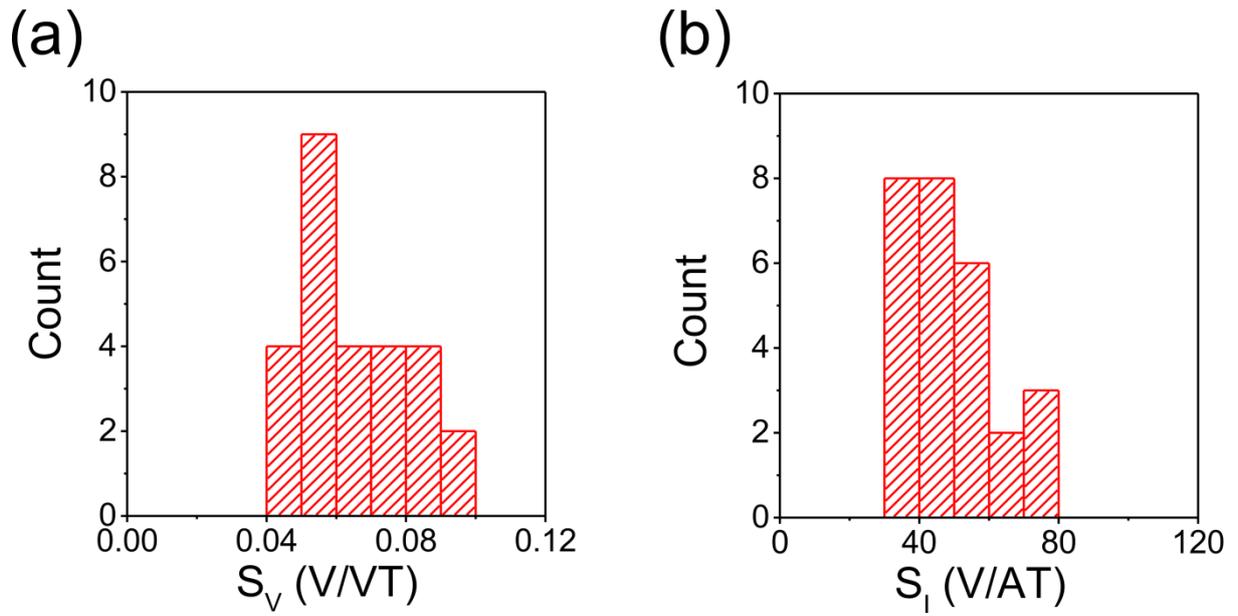

Figure 3. Histogram of the (a) voltage normalized sensitivity and (b) current normalized sensitivity for different Hall sensors on the same sample chip.

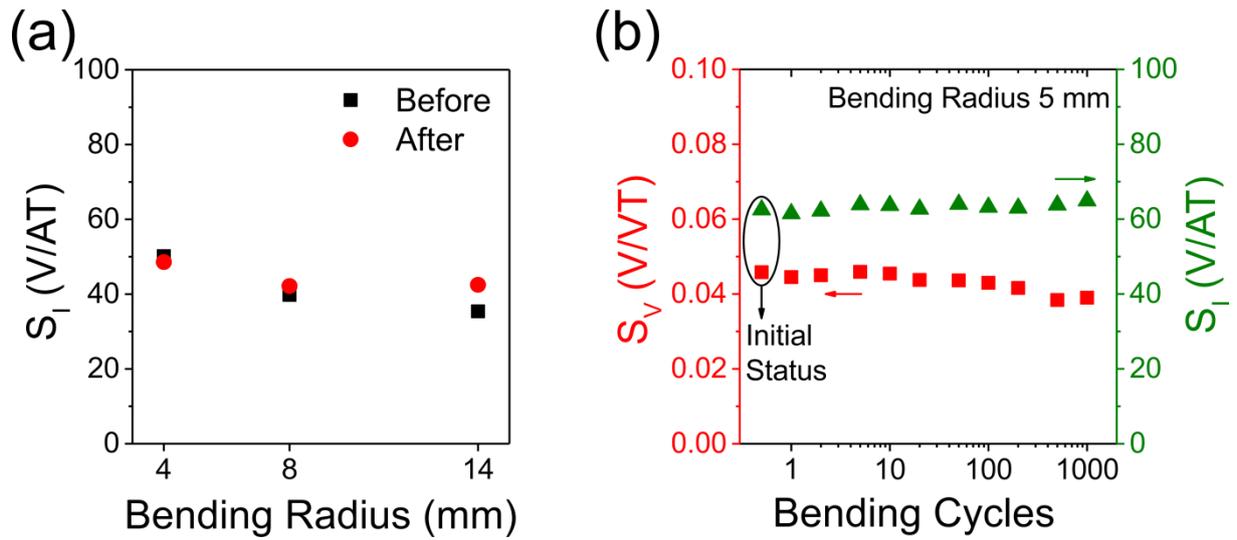

Figure 4. The bending test of the flexible Hall sensor. (a) The current normalized sensitivity for the samples before and after one time bending to specific banding radii. The graphene Hall sensors were always located at the outer surface, so that tensile strain is applied. The measurements are carried out in flat status. (b) The bending cycles measurement for the sensitivity of the Hall sensor. The sensors are bent with a radius of 5 mm up to 1000 times.

Table I. Comparison of the sensitivity parameters from different types of Hall sensors. The numbers in brackets are the mean values measured for the different devices on the substrate in this work.

|  | Substrate | $S_I$ (V/AT) | $S_V$ (V/VT) |
|---|---|---|---|
| Graphene in this work | Flexible | 75 (49) | 0.093 (0.065) |
| Bismuth[17] | Flexible | 2.3 | N/A |
| Si[3,5] | Rigid | 100 | 0.1 |
| Graphene[23] | Rigid | 5700 | 3 |

TOC Entry

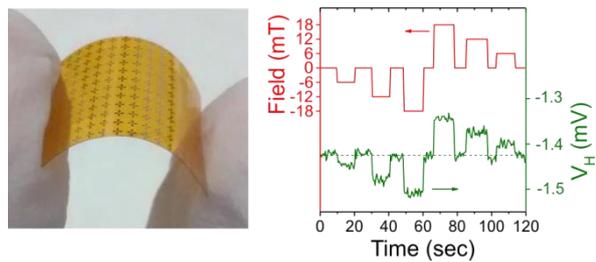

Fully flexible and ultrathin Hall sensor based on graphene with high sensitivity for magnetic field detection and stability under bending